\documentclass[aps,pre,superscriptaddress, amsmath, amssymb, showpacs, twocolumn]{revtex4-1}

\newcommand{\vc}{\mathbf}
\usepackage{graphicx}
\usepackage{color}

\begin{document}

\title{Friction Force in Strongly Magnetized Plasmas }

\author{David J. Bernstein}
\affiliation{Department of Physics and Astronomy, University of Iowa, Iowa City, Iowa 52242, USA}
\author{Trevor Lafleur}
\affiliation{PlasmaPotential-Physics Consulting and Research, Canberra, ACT 2601, Australia}
\author{J\'er\^ome Daligault}
\affiliation{Los Alamos National Laboratory, Los Alamos, New Mexico 87545, USA}
\author{Scott D. Baalrud}
\email[]{scott-baalrud@uiowa.edu}
\affiliation{Department of Physics and Astronomy, University of Iowa, Iowa City, Iowa 52242, USA}

\date{\today}

\begin{abstract}

A charged particle moving through a plasma experiences a friction force that commonly acts antiparallel to its velocity. 
It was recently predicted that in strongly magnetized plasmas, in which the plasma particle gyro-frequency exceeds the plasma frequency, the friction also includes a transverse component that is perpendicular to both the velocity and Lorentz force. 
Here, this prediction is confirmed using molecular dynamics simulations, and it is shown that the relative magnitude of the transverse component increases with plasma coupling strength. 
This result influences single particle motion and macroscopic transport in strongly magnetized plasmas found in a broad range of applications. 
\end{abstract}

\maketitle

Friction influences the dynamics of projectiles as they travel through a medium. 
It also determines how interactions at microscopic scales influence the macroscopic rate of particle, momentum, and energy transfer. 
In plasmas, as in other media, friction is commonly thought to act antiparallel to the velocity of a projectile, here considered to be a single charged particle. 
Recent work has predicted that a qualitatively different effect arises in strongly magnetized plasmas, which are characterized by the property that the gyro-frequency of the plasma species ($\omega_c$) responsible for slowing the projectile significantly exceeds its plasma frequency ($\omega_p$) \cite{Lafleur_Baalrud_2019}. 
The predicted effect is a component of the friction force that is perpendicular to both the projectile velocity and Lorentz force vectors.
This transverse force depends on the orientation of the velocity vector with respect to the magnetic field, and causes the total friction force to shift with respect to the particle's velocity vector in the plane of the velocity and magnetic field.

Here, the prediction of a transverse component of friction is confirmed using first-principles molecular dynamics simulations. 
Plasmas with a strongly magnetized component are found in many instances, some of which are summarized in Table \ref{tab:tab1}. 
For example, electrons and positrons in antimatter traps and non-neutral plasmas are strongly magnetized \cite{Surko_Fajan,nonneutral_PRL_1977,nonneutral_PRL_1980,nonneutral_PhysFluids_1980}. 
Key steps in these experiments include slowing and cooling antiprotons on strongly magnetized electrons via friction, and mixing antiprotons with strongly magnetized positrons to create anti-hydrogen. 
The transverse force will alter the trajectory of the antiprotons and may influence the confinement rate as it increases the gyro-radius of particles moving faster than the thermal speed of the background plasma \cite{Lafleur_Baalrud_2019}. 
Magnetized ultra-cold neutral plasmas are a new experimental platform in which electrons can access the strongly magnetized regime at modest applied magnetic field strengths because of the low plasma density and temperature \cite{Ultracold_2}. 
The transverse friction force will influence the dynamics of ions in these experiments, which in turn influences the macroscopic expansion of the plasma. 
In astrophysics, the atmosphere of neutron stars and electrons in the magnetosphere of Jupiter, and likely many exoplanets, are strongly magnetized \cite{neutron_star_1,Jupiter}. 
The transverse friction force will influence the rate of particle and momentum transport in these systems. 
Finally, we note that even in magnetic confinement fusion experiments, the electrons can be in a modestly strongly magnetized regime \cite{ITER_parameters}. 
The transverse force may influence the trajectory of fusion products, or runaway electrons, in these experiments \cite{ITER_ion_heating,runaway_electrons}. 

\begin{table*}
\begin{tabular}{ccccccc}
\hline
\hline
System & $n_e$ (cm$^{-3}$) & $k_BT_e$ (eV) & $B$ (T) & $\beta_e$ & $\Gamma_e$ & Ref. \\
\hline
antimatter traps& $ 10^8 $ & $10^{-3}$ & 1 & 300 & 0.1 &  \cite{Surko_Fajan}\\
neutron star atmospheres & $10^{24}$ & $100$ & $10^8$ & 300 & 0.3 & \cite{neutron_star_1}\\
Jupiter magnetosphere  & $10^3$& 100 & $4 \times 10^{-4}$ & 30 & $10^{-8}$ & \cite{Jupiter}\\
ultra-cold neutral plasmas & $10^{7}$ & $4 \times 10^{-4}$ & 0.01 & 10 & 0.1 &  \cite{Ultracold_2} \\
magnetic confinement fusion & $10^{14}$ &$ 2 \times 10^4$ & 5 & 2 & $10^{-7}$ & \cite{ITER_parameters}\\
\hline
\hline
 \end{tabular}
\caption{A sample of systems for which the electrons (or positrons) are strongly magnetized. Columns list electron density ($n_e$), electron temperature ($k_BT_e$), external field strength ($B$), magnetization parameter ($\beta_e = \omega_{c_e}/\omega_{p_e}$), and electron coupling strength ($\Gamma_e$).
For more antimatter trap parameters, see Table I in \cite{Surko_Fajan}.}
\label{tab:tab1}
\end{table*}

Accounting for the transverse component, the total friction force on a test charge can be expressed as \cite{Lafleur_Baalrud_2019}
\begin{equation}
\label{eq:friction}
\vc{F} = F_v \hat{\vc{V}} + F_\times \hat{\vc{V}} \times \hat{\vc{n}}
\end{equation}
in which $\hat{\vc{n}} = (\hat{\vc{V}} \times \hat{\vc{B}})/\sin \theta$, where $\theta$ is the angle between the velocity $\vc{V}$ and magnetic field $\vc{B}$ ($\hat{\vc{V}} = \vc{V}/|\vc{V}|$ and $\hat{\vc{B}} = \vc{B}/|\vc{B}|$). 
In a weakly magnetized plasma, the friction force can be computed using traditional plasma kinetic equations \cite{Zwick_Review}, such as the Landau-Boltzmann equation \cite{Landau} or the Lenard-Balescu equation \cite{Lenard,Balescu}. 
These have collision operators that do not depend on the magnetic field because the gyro-radius is assumed to be much larger than the Debye scale over which particle interactions occur. 
These lead to the prediction that $F_\times = 0$, and the friction is entirely determined by $F_v$ ($-F_v$ is commonly referred to as the stopping power \cite{Zwick_Review}). 
A significant body of work has been developed to extend both the Landau-Boltzmann and Lenard-Balescu approaches to the strongly magnetized regime \cite{ONeil_1980,Rostoker_1960,Montgomery,Mag_dEdx_Book}. 
These have shown that the stopping power ($-F_v$) is significantly altered by the magnetic field in the strongly magnetized regime \cite{Mag_dEdx_Book}. 
However, they considered only $F_v$ rather than the full friction vector. 
It was only recently suggested that there is a second transverse component of the friction force that is perpendicular to the velocity vector \cite{Lafleur_Baalrud_2019}. 
This prediction was made using a linear response approximation, which has never been validated in the strongly magnetized regime. 
Validation is a critical step to acceptance of this predicted physical effect.

\begin{figure*}
\centering
\includegraphics[width=14cm]{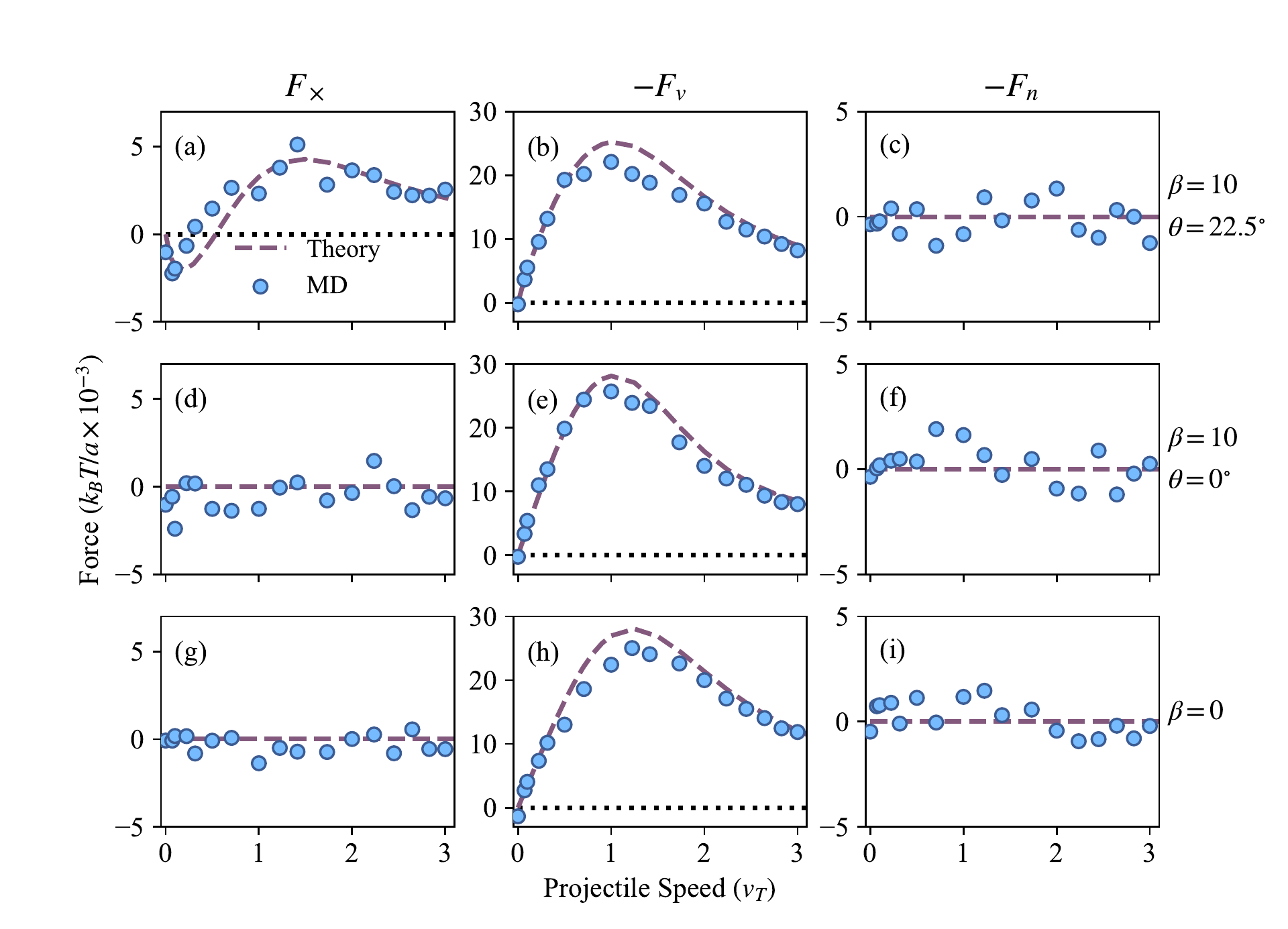}
\caption{A comparison between theoretical predictions and simulation results for the transverse force ($F_\times$) [panels (a), (d) and (g)], stopping force ($-F_v$) [(b), (e), and (h)], and force in the Lorentz force direction ($-F_{n}$) [(c), (f), and (i)].
Parameters $\beta$ and $\theta$ for each row are indicated.
The black dotted line marks 0.
$\Gamma = 0.1$ for these simulations.}
\label{fig:force}
\end{figure*}

The goal of this work is to verify the existence of the transverse force using molecular dynamics (MD) simulations.
These simulations provide a means of validation because they are based only on first-principles. 
As long as numerical convergence is obtained, they make no approximation other than the validity of Newton's equations of motion, the Coulomb force law, and the Lorentz force operator in the presence of an external magnetic field.
Friction is an average force.
Therefore, many projectiles traveling through a plasma were simulated for a short period of time, and the average force exerted on them by the plasma was computed.
By analyzing the force parallel and perpendicular to the projectile velocity, the presence of the transverse component was verified.

The background plasma was modeled using the magnetized one-component plasma (OCP) model, which consists of a single species of charged particles interacting via the Coulomb force, as well as the Lorentz force associated with the external field, in which the neutralizing background is non-interacting \cite{Baalrud_Daligault_MagPhases,Ott_Bonitz_PRL,BausHansen}.
It is fully characterized by two parameters.
The first is the Coulomb coupling parameter $\Gamma = (q^2/ a)/(k_BT)$, which is the ratio of the potential energy at the average inter-particle spacing ($a = (3/4\pi n)^{1/3}$ where $n$ is the density) to the average kinetic energy ($k_BT$) \cite{Ichimaru,Footnote_Units}.
The second is the magnetization parameter $\beta = \omega_c/\omega_p$, where $\omega_c = q|\vc{B}|/cm$ is the gyro-frequency and $\omega_p = \sqrt{4 \pi nq^2/m}$ is the plasma frequency.
When $\beta \gg 1$, the plasma is considered strongly magnetized \cite{Baalrud_Daligault_MagPhases}.
When this condition is met, the particle gyration occurs on the same time and length scales as microscopic collisions \cite{Baalrud_Daligault_MagPhases}.
Although the OCP is a model system, it is well adapted to validate the friction force because a particle of a given energy predominately interacts only with the plasma species of a similar kinetic energy. 
For instance, these simulations accurately represent a fast ion slowing on the electrons of a background neutral plasma.

Here, the projectile has a mass of $M = 1000m$ and charge $Q = q$.
Its gyro-frequency is therefore 1000 times smaller than that of the plasma particles.
The simulation duration of a few plasma periods is sufficiently short that the gyro-motion of the projectile is negligible. 
The magnetic field only indirectly affects the projectile via the friction force exerted by the magnetized background.

The simulations were conducted using the code described in ~\cite{Code}, which utilizes the particle-particle-particle-mesh method \cite{MDbook}.
They evolved $5 \times 10^4$ particles in a cubic periodic domain, corresponding to a domain length of $59a$.
First, an unmagnetized plasma was equilibrated for $3.05 \times 10^4 \omega_p^{-1}$ with a velocity scaling thermostat \cite{MDbook}.
The magnetic field was not included during the equilibration phase because the relaxation to equilibrium is faster without it, and the magnetic field does not influence the equilibrium state (the Bohr-van Leeuwen theorem \cite{Pathria}). 
Particle velocities and positions were recorded every $1 \omega_p^{-1}$ after the first $500 \omega_p^{-1}$, yielding $3 \times 10^{4}$ independent particle configurations.
Time was discretized into timesteps of $0.001 \omega_p^{-1}$, which is much smaller than the timescale at which close collisions and  particle gyrations occur.

From each of the $3 \times 10^4$ configurations collected, an individual simulation was conducted consisting of three steps.
1) The thermostat was turned off and a magnetic field oriented along the $+z$ direction was imposed with magnetic field strength corresponding to either $\beta = 0$ or 10.
2) A projectile was introduced and launched in the $x-z$ plane with speed $V_0$ and angle $\theta$ between the velocity and magnetic field, where $V_0$ is in units of the plasma thermal speed $v_T = \sqrt{2k_BT/m}$.
There is a short transient period in which the plasma responds to the abrupt introduction of the projectile which lasts for about $1-2 \omega_p^{-1}$.
After this period, the projectile energy loss is steady \cite{Bernstein_Baalrud_Daligault_2019}.
To remove effects from the transient period, the projectile's momentum was held constant for $2 \omega_p^{-1}$.
Particle positions were recorded at the end of this step, and used for potential distribution calculations.
3) After the transient period passed, the projectile's momentum was no longer held constant, letting the projectile fully interact with the plasma.
The total force induced by the plasma particles on the projectile was tracked over time.
The forces in cartesian coordinates ($F_x$, $F_y$, and $F_z$) were recorded every $0.01 \omega_p^{-1}$ (every 10 timesteps) for a total time of $1 \omega_p^{-1}$.
A time average of the forces was calculated for each simulation.
The results of the time averages were then averaged over the $3 \times 10^4$ simulations, and used in $F_\times = F_x \cos \theta - F_z \sin \theta$, $F_v = F_x \sin \theta + F_z \cos \theta$, and $-F_{n} = F_y$ to yield the final results for the transverse and stopping forces, and the force perpendicular to the plane defined by the velocity and magnetic field (Fig.~\ref{fig:force}).
The stopping power is $-F_v$, which describes the average energy loss with respect to distance, and the forces are displayed in units of $k_BT/a$ \cite{Zwick_Review}.

The main result of this study is shown in Fig.~\ref{fig:force}(a): 
A transverse force on the projectile is present in the simulations when $\beta = 10$ and $\theta = 22.5^{\circ}$ [Fig.~\ref{fig:force}(a)], in good agreement with the theoretical prediction (the transverse force is predicted to be largest when $\theta \approx 22.5^{\circ}$ \cite{Lafleur_Baalrud_2019}).
MD simulation results are shown as data points, and theoretical predictions from \cite{Lafleur_Baalrud_2019} are shown as dashed lines.
No transverse force was observed when $\beta = 10$ and $ \theta = 0^{\circ}$, or when $\beta = 0$ [Figs.~\ref{fig:force}(d) and (g)], which is also consistent with the theory.
The simulations and theoretical predictions quantitatively agree. 
The theory is based on a linear response approximation, whereas the MD simulations are first-principles non-linear computations. 
They provide validation of the prediction of the transverse friction force.

The predictions and MD results for the stopping forces also agree well [Figs.~\ref{fig:force}(b), (e), and (h)].
The MD data is slightly lower than the predictions around the Bragg peak, which was previously seen in simulations of unmagnetized plasmas at the same coupling strength \cite{Bernstein_Baalrud_Daligault_2019,Footnote_1}.
The friction force in the direction perpendicular to the velocity-magnetic field plane (the $\hat{\vc{n}}-$direction) is also shown [Figs.~\ref{fig:force}(c), (f), and (i)]. 
Theory predicts $-F_{n} = 0$, which the simulation results fluctuate about with no discernible signal.

\begin{figure}
\centering
\includegraphics[width=8.6cm]{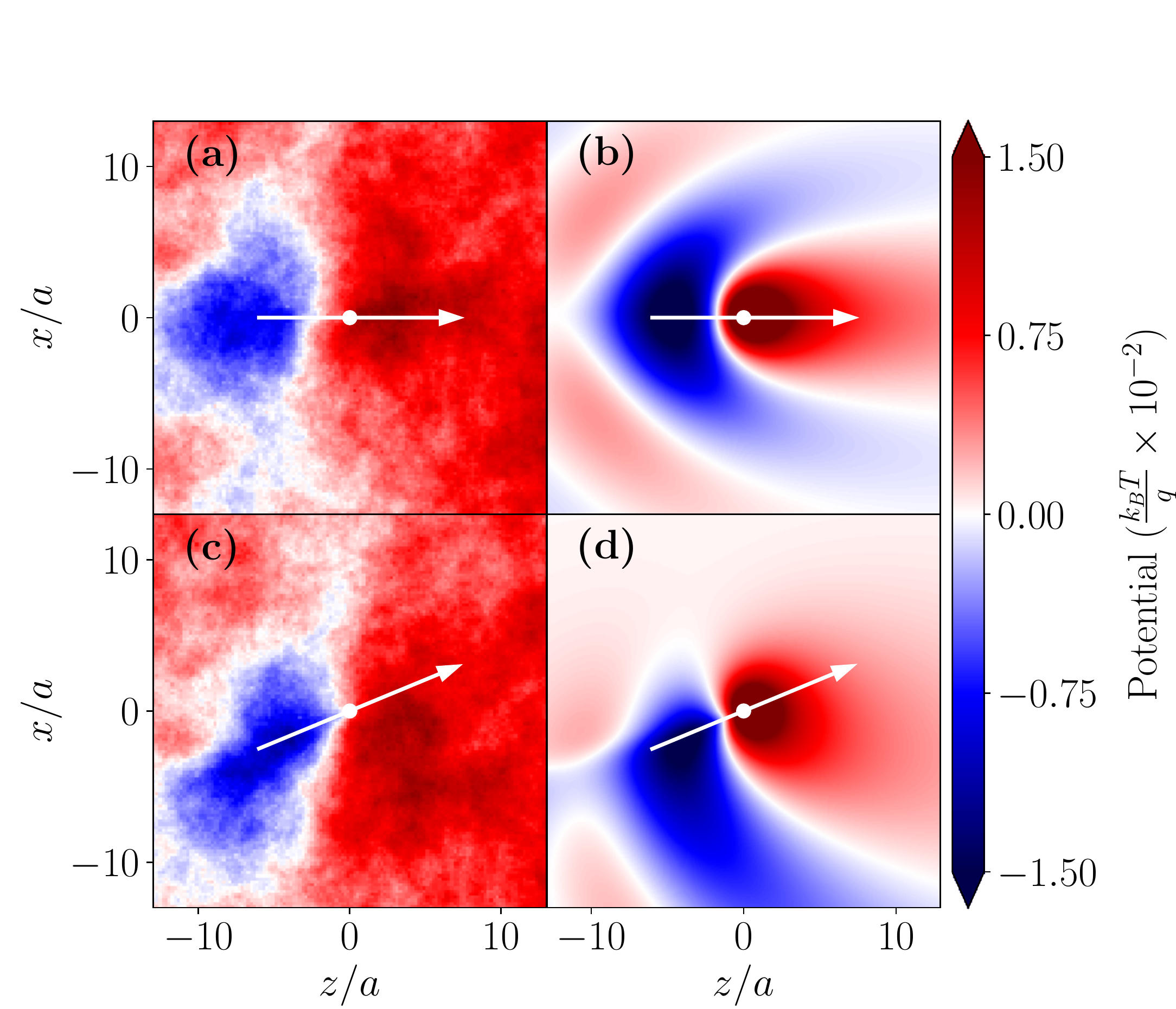}
\caption{Electrostatic potential maps (wakes) from MD simulation are shown in panels (a) and (c) ($\beta = 10$), and from linear response theory in the large $\beta$ limit \cite{Lafleur_Baalrud_2019} are shown in panels (b) and (d).
In panels (a) and (b), $\theta = 0^{\circ}$.
In panels (c) and (d), $\theta = 22.5^{\circ}$.
The positions of the projectiles are marked with white dots, and the arrows show the velocity orientation and direction ($V_0=2v_T$).
}
\label{fig:potentials}
\end{figure}

In linear response theory, the friction force is computed from the electric field induced by the electrostatic potential wake that forms about the projectile \cite{Ichimaru}.
The wake is distorted in the presence of a magnetic field \cite{Shukla_Salimullah_1996,Darian_Miloch_2019,Piel_Greiner_2018,Ware_Wiley_1993,Joost_Ludwig_2014}, and can be asymmetric about $\vc{V}_0$ if $\theta$ is not $0^{\circ}$ or  $90^{\circ}$ \cite{Lafleur_Baalrud_2019}.
The transverse force arises from the induced electric field associated with this asymmetry \cite{Lafleur_Baalrud_2019}.
To see this, wakes were calculated with the particle positions recorded at the end of step 2 (Fig.~\ref{fig:potentials}).

To calculate the wakes, a two-dimensional grid in the $x-z$ plane consisting of $250 \times 250$ points was established for each simulation centered at the projectile's final position.
The Coulomb potential at each grid point was calculated from particles within a sphere of radius $52a$ about the grid point.
To account for quasi-neutrality, the OCP model includes an immobile, non-interacting, neutralizing background \cite{BausHansen}.
The constant potential from this theoretical background was subtracted at each grid point in order to compare the MD results with theory.
The potential values on the grids were then averaged across the simulations, yielding the results shown in Fig.~\ref{fig:potentials}(a) and (c).

When $\theta = 0^{\circ}$, the wakes from both the simulations and theory are symmetric about the $z-$ axis [Fig.~\ref{fig:potentials}(a) and (b)].
Since the friction force can be attributed to the electric field induced by the charge distribution, a wake that is symmetric about the velocity vector implies that $F_\times = 0$.
When $\theta = 22.5^{\circ}$, an asymmetry about $\vc{V}_0$ is observed in the wakes in both the simulations and predictions [Figs.~\ref{fig:potentials}(c) and (d)]; suggesting that $F_\times$ is non-zero, as is confirmed in Fig.~\ref{fig:force}(a).

\begin{figure}
\centering
\includegraphics[width=8.6cm]{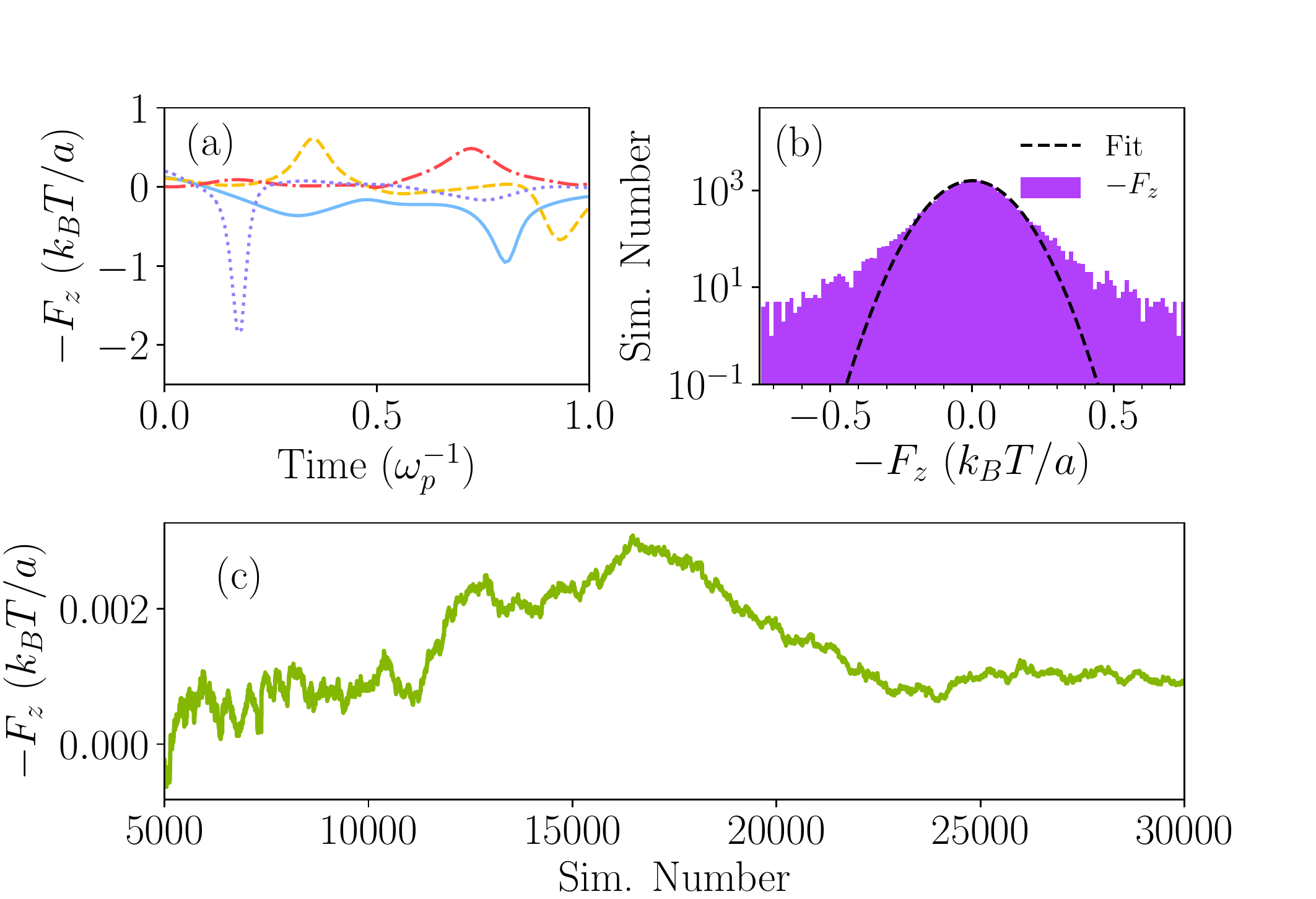}
\caption{(a) Individual force time series (along the $z$-direction) for four different simulations (with $\beta = 10$, $\theta = 0^{\circ}$, and $V_0 = 2v_T$).
(b) Distribution of $3 \times 10^4$ time-averaged force measurements compared with a best fit Gaussian. 
(c) Cumulative average of the time-averaged force measurements with simulation number.
$\Gamma = 0.1$ for all panels.}
\label{fig:stats}
\end{figure}

A large number of simulations was needed to verify the presence of the transverse force by reducing the relatively large statistical noise.
A large range of time-averaged force values was observed [examples of the force time series are given in Fig.~\ref{fig:stats}(a)], forming a broad distribution [an example of this is given in Fig.~\ref{fig:stats}(b)].
The histogram in Fig.~\ref{fig:stats}(b) does not show the full extent of one of the tails, which extends to $-F_z \approx 4 k_BT/a$.
By comparing the distribution of time averages to a best fit Gaussian [Fig.~\ref{fig:stats}(b)], one can see that the distribution is skewed and the tails are heavily populated.
A large number of simulations was needed to populate the distribution and accurately calculate the mean.
This is shown in Fig.~\ref{fig:stats}(c), where the cumulative average is displayed.
The error bars for each data point have a size of one standard deviation of the mean, which are smaller than the data markers in Fig.~\ref{fig:force}.
There are both small fluctuations that occur over small numbers of simulations, and drifts that occur over large numbers of simulations [Fig.~\ref{fig:stats}(c)].
It is the latter of these that could be contributing to the fluctuations in the simulation results.

\begin{figure}
\centering
\includegraphics[width=8.5cm]{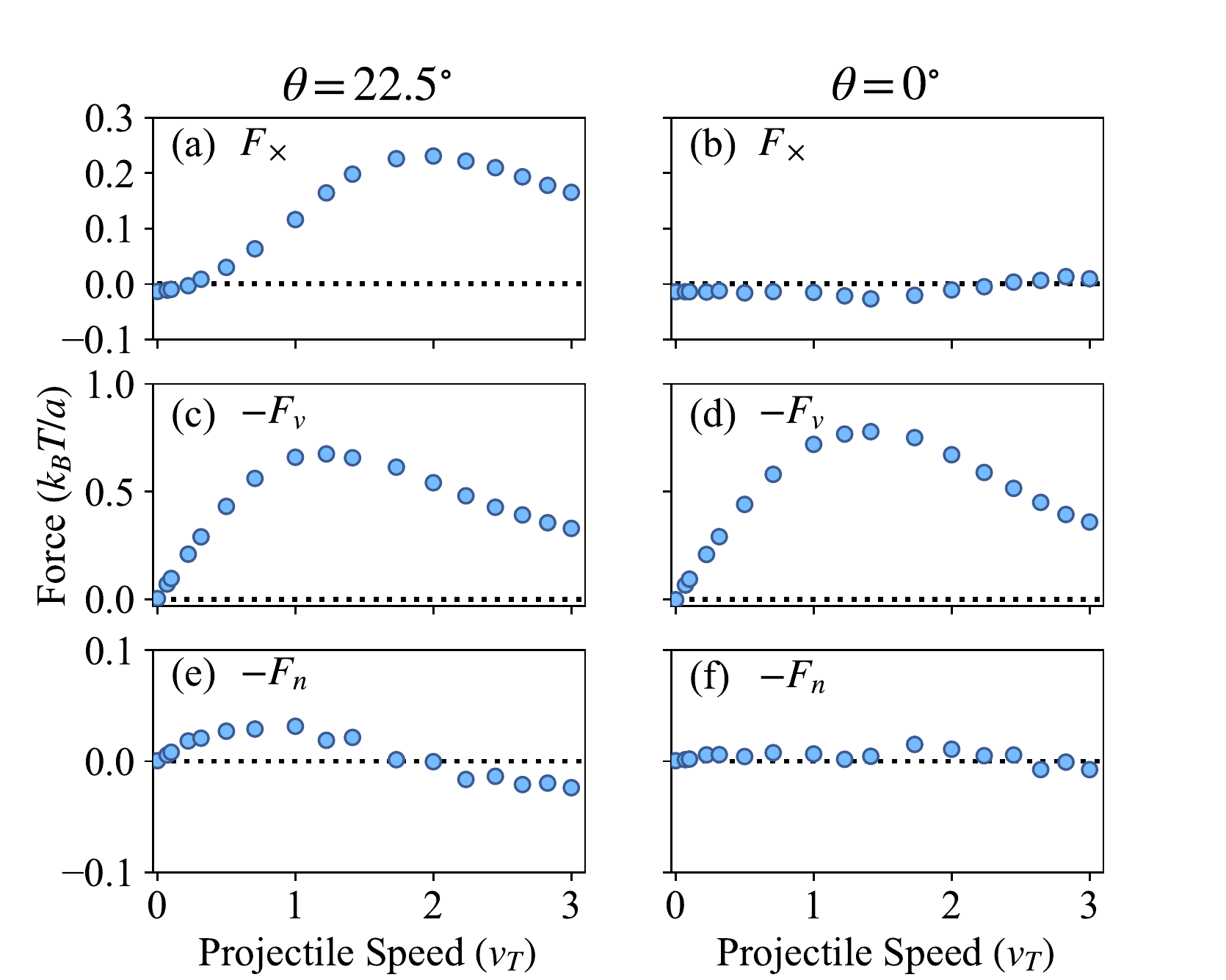}
\caption{Simulation results (blue dots) for forces when $\Gamma =1$ and $\beta = 10$. 
The dotted line marks 0.
Panels (a) and (b) show $F_\times$, (c) and (d) show $-F_v$, and (e) and (f) show $-F_{n}$.
$1 \times 10^4$ particles were used for these simulations (less than the $\Gamma = 0.1$ simulations).
}
\label{fig:Gamma1}
\end{figure}

Some strongly magnetized plasmas, such as non-neutral plasmas, exist at conditions of moderate to strong Coulomb coupling. 
Previous work has suggested that for a fixed value of $\beta$ the magnetic field more strongly influences transport as the coupling strength increases from a weak to moderate value \cite{Baalrud_Daligault_MagPhases}; e.g., if $\Gamma$ increases from 0.1 to 1 at $\beta = 10$. 
The reason is that the gyro-radius transitions from being larger than the distance of closest approach in a binary collision, to being shorter than it. 
Although the theory from \cite{Lafleur_Baalrud_2019} only applies to weakly coupled plasmas, it also predicts that the transverse force becomes a larger component of the total friction force as this regime is approached. 
Figure~\ref{fig:Gamma1} shows that, as expected, the ratio of the maximum transverse force compared to the maximum stopping force in the $\Gamma = 1$ case is larger than in the $\Gamma = 0.1$ case [compare Fig.~\ref{fig:Gamma1}(a) with Fig.~\ref{fig:Gamma1}(c), and compare Fig.~\ref{fig:force}(a) with Fig.~\ref{fig:force}(b)].
Likewise, the stopping power exhibits previously predicted behavior as the angle increases, i.e. the maximal stopping force shifts to lower speeds and the magnitude decreases \cite{Mag_dEdx_Book}.
In addition, a statistically significant component of the friction force in the Lorentz force direction ($-F_n$) is observed for $\theta=22.5^\circ$. 
This is not predicted by the linear response theory, indicating that a non-linear effect becomes important at these moderately coupled and strongly magnetized conditions. 
Molecular dynamics simulations are first principles, whereas the linear response theory is based upon a weak interaction approximation. 
The mechanism responsible for this observation will be studied in greater detail in future work.

In conclusion, this work has confirmed a predicted transverse friction force in strongly magnetized plasmas using first-principles MD simulations.
This is associated with gyro-motion at the microscopic scale of collisions, and significantly alters the average trajectory on macroscopic scales \cite{Lafleur_Baalrud_2019}.
This could affect how well particles are contained in experiments that rely on magnetic confinement, such as anti-matter traps, non-neutral plasmas, and fusion experiments.
The existence of the transverse force exemplifies shortcomings of traditional plasma kinetic theory.
Kinetic theories that are adapted to accurately account for strong magnetization can be tested by calculating this force.

This material is based upon work supported by the U.S. Department of Energy, Office of Science, Office of Fusion Energy Sciences under Award Number DE-SC0016159 under Award Number DE-NA0003868 and by the National Science Foundation under Grant No.~PHY-1453736. 
It used the Extreme Science and Engineering Discovery Environment (XSEDE), which is supported by NSF Grant No. ACI-1053575, under Project Award No. PHYS-150018. 
The work of JD was performed under the auspices of the U.S. Department of Energy under Contract No. 89233218CNA000001 and was supported in part by the U.S. Department of Energy LDRD program, Grant No. 20170073DR, at Los Alamos National Laboratory.


\bibliography{refs.bib}

\begin{thebibliography}{99}
\bibitem{Footnote_Units}
This work is done in Gaussian centimeter-gram-second units.
\end{thebibliography}


\end{document}